\documentclass[fleqn,twocolumn]{aastex631}
\usepackage{amsmath}

\newcommand\beq{\begin{equation}}
\newcommand\eeq{\end{equation}}
\newcommand{\beqa}{\begin{eqnarray}}
\newcommand{\eeqa}{\end{eqnarray}}
\newcolumntype{M}[1]{>{\centering \arraybackslash}m{\if!#1!3cm\else#1\fi}}
\newcommand\beqn{\begin{eqnarray}}
\newcommand\eeqn{\end{eqnarray}}
\newcommand\be{\begin{eqnarray}}
\newcommand\ee{\end{eqnarray}}

\newcommand{\newsuggestion}[1]

\usepackage{ae,aecompl}
\usepackage[utf8x]{inputenc}

\usepackage{url}
\usepackage{multirow}

\shorttitle{HI-tSZ cross-correlation}
\shortauthors{Ibitoye et al.}

\graphicspath{{./}{figures/}}

\begin{document}

\title{HI Intensity Mapping cross-correlation with thermal SZ fluctuations: forecasted cosmological parameters estimation for FAST and Planck}
\author[0000-0002-0966-8598]{Ayodeji Ibitoye}
\thanks{Corresponding author: A.Ibitoye,\url{astro.ayodejiibitoye@gmail.com}}

\affiliation{National Astronomical Observatories, Chinese Academy of Sciences, \\20A Datun Road, Chaoyang District, Beijing 100101, P. R. China}
\affiliation{Centre for Space Research, North-West University, Potchefstroom 2520, South Africa}
 \affiliation{NAOC-UKZN Computational Astrophysics Centre (NUCAC), University of KwaZulu-Natal, Durban, 4000, South Africa}
\affiliation{Department of Physics and Electronics, Adekunle Ajasin University, P. M. B. 001, Akungba-Akoko, Ondo State, Nigeria}

\author{Furen Deng}
\affiliation{National Astronomical Observatories, Chinese Academy of Sciences, 
\\20A Datun Road, Chaoyang District, Beijing 100101, P. R. China}

\author{Yichao Li}
\affiliation{Key Laboratory of Cosmology and Astrophysics (Liaoning) \& College of Sciences, \\Northeastern University, Shenyang 110819, China}

\author[0000-0001-8108-0986]{Yin-Zhe Ma}
\affiliation{Department of Physics, Stellenbosch University, Matieland, Western Cape, 7602, South Africa}

\author[0000-0001-6475-8863]{Xuelei Chen}
\thanks{Corresponding author: X. Chen, \url{xuelei@cosmology.bao.ac.cn}}
\affiliation{National Astronomical Observatories, Chinese Academy of Sciences, 
\\20A Datun Road, Chaoyang District, Beijing 100101, P. R. China}
\affiliation{Key Laboratory of Cosmology and Astrophysics (Liaoning) \& College of Sciences, \\Northeastern University, Shenyang 110819, China}
\affiliation{Key Laboratory of Radio Astronomy and Technology, Chinese Academy of Sciences, Beijing 100101, China}

\begin{abstract}
The 21 cm emission from neutral hydrogen surveys holds great potential as a valuable method for exploring the large-scale structure of the Universe. In this paper, we forecast for the cross-correlation between the Thermal Sunyaev-Zel'dovich (SZ) fluctuations as probed by the Planck satellite, and fluctuations in the HI brightness temperature as probed by the ground-based Five-hundred-meter Aperture Spherical Telescope (FAST), to trace the connection between galaxy clusters and the HI large-scale structure. Assuming that the measurement is limited by instrumental noise rather than by foreground, we estimate the potential detectability of the cross-correlation signal and their improvement in the measurement of the HI cosmic density, the hydrostatic mass bias parameter, and the universal pressure profile (UPP) parameters.
We obtain a constraint on the cosmic neutral hydrogen density parameter significantly to $\sigma(\Omega_{\rm HI}) = 1.0 \times 10^{-6}$. We also find that the average halo masses contributing to the ${{\rm HI}-y}$ cross-power spectrum in the one-halo regime is $\sim 1.5\times 10^{14} M_{\odot}$. Our results also show that the HI-SZ cross-correlation has great potential to probe the distribution of neutral hydrogen (HI) within halos at low redshift.
\end{abstract}



\section{Introduction } 
\label{sec:introduction} 
Taking into account cosmic variance, achieving greater precision in cosmological measurements requires surveying more extensive volumes of the Universe. The ultimate aim is to map the entirety of the observable universe. Neutral hydrogen (HI) is the most appropriate tracer for this, as it pervades space from the recombination era to the present day. The 21 cm line of HI is redshifted, offering a method to measure cosmic distance and allowing for the reconstruction of the three-dimensional density field of matter across various redshifts and scales.

At redshifts below $z = 0.1$, the study of HI involves galaxy surveys of 21 cm emission \citep{Zwaan2005,Martin10}. More recently, 
the abundance and distribution of neutral hydrogen are inferred from a catalog of 41741 extragalactic HI sources from the FAST all sky HI survey \citep{Zhang24}. However, at higher redshift, current radio observations of individual galaxies are still very few, and the existing constraints from studies conducted by \citet{Rao06,Lah07,Meiring11} have significant uncertainties. Though, at $z>2.2$, the HI can be probed by ground-based optical observation using its Lyman-$\alpha$ absorption \citep{Prochaska09, Bird17,Ho21}.

Although the HI line is not bright enough for individual galaxies to be observed at $z>0.1$, there is still the possibility of conducting three-dimensional (3D) intensity mapping. This involves studying the large-scale structure (LSS) directly by detecting the combined emission from numerous galaxies that occupy sizeable voxels of approximately $1000 \,{\rm Mpc^3}$ \citep{Chang_Tzu08,Loeb08,Mao08,Seo10,Ansari12}. Instead of cataloging individual galaxies, this method allows for the examination of the overall emission from multiple galaxies. By utilizing such large voxels, telescopes (existing and upcoming) can effectively survey HI intensity. Several telescopes and surveys are in preparation to do this, including CHIME, Tianlai, FAST, BINGO, HIRAX  \citep{CHIME_Chapter6, Chen2012,Xu15,Perdereau22,Hu_2020,FAST_drift_scan,BINGO,HIRAX} and of course the Square Kilometer Array pathfinders; MeerKAT , and ASKAP \citep{MeerKAT_Chapter6,ASKAP}.   This approach is ideal for cosmological surveys, as shown in \citep{Jin21,Zhang21,Wu23a,Wu23,FAST_drift_scan}. 
 
Until the recent first report of HI auto-power spectrum signal detection on ${\rm Mpc}$ scale using MeerKAT radio telescope data by \citet{Paul23}, the detection of the HI signal has relied on the use of cross-correlation techniques. \citet{first_HI_science} published the first cross-correlation detection of the underlying HI ($\simeq 4 \sigma$) signal between LSS and HI intensity maps at, $z \sim 1$ using data from DEEP2 galaxy survey and the GBT. A significant improvement on this was made 
by \citet{Masui13} with an increased integration time (15hr) of the GBT data in cross-correlation with WiggleZ Dark Energy Survey \citep{Drinkwater10}. Thereafter, the first upper bound constraints on the neutral hydrogen fluctuations at $z \sim  0.8$ from the auto-power spectrum of the HI 21 cm intensity fluctuations (using the increased GBT integration time 190hr) were reported by \citet{Switzer13}. 

Within a redshift range of $ 0.6 < z < 1$, the cross-correlation data constrained the product of the neutral hydrogen (HI) fraction, the galaxy–hydrogen correlation coefficient ($r$), and the HI bias parameter ($b_{\rm HI}$), as $\Omega_{\rm HI} b_{\rm HI} r_{\rm HI}= [0.43 \pm 0.07\, {\rm (stat.)} \pm 0.04 \,{\rm (sys.)}] \times 10^{-3}$. Other LSS cross-correlations with HI intensity maps, however, have provided useful constraints on the underlying HI signal. For example, \citet{Anderson2018} reported the cross-correlation analysis done using the 21${\rm cm}$ intensity maps from Parkes Telescope data which span a redshift of $0.057< z <0.098$ with the Two-degree-Field Galaxy Redshift Survey(2dFGRS), and obtained a significance detection of $14.8 \sigma$, and inferred that it is possible to observe a more correlated signal of HI with blue-type galaxies. In the same light, \citet{LauraWolz22} placed a constraint on $\Omega_{\rm HI} b_{\rm HI} r_{\rm HI}$ by studying the cross-correlation of the HI intensity mapping observation data with three optical galaxy samples: the Luminous Red Galaxy (LRG) and Emission Line Galaxy (ELG) samples from the eBOSS survey, and the WiggleZ Dark Energy Survey sample. At an effective scale of $ k_{\rm eff} = 0.31 {\rm h} \,{\rm Mpc}^{-1}$, the cross-correlation signal was used to obtain $\Omega_{\rm HI} b_{\rm HI} r_{\rm HI, wig}= [0.58 \pm 0.09\, {\rm (stat)} \pm 0.05 \,{\rm (sys)}] \times 10^{-3}$,  for GBT-WiggleZ, $\Omega_{\rm HI} b_{\rm HI} r_{\rm HI,ELG} = [0.40 \pm 0.09 \,{\rm (stat)} \pm 0.04 \,{\rm (sys)}] \times 10^{-3}$ for GBT-ELG, and $\Omega_{\rm HI} b_{\rm HI} r_{\rm HI, LRG} = [0.35 \pm 0.08\, {\rm (stat)} \pm 0.03 \,{\rm (sys)}] \times 10^{-3 }$ for GBT-LRG, at $z \simeq 0.8$. Recently, \citet{Cunnington23} presented a $7.7 \sigma$ direct detection of correlated clustering between the MeerKAT radio intensity maps and galaxies from the WiggleZ Dark Energy survey. 

The HI cosmic density has been studied previously both at low and high redshift. At low redshift, the direct measurement of $\Omega_{\rm HI}$ is possible by calculating the total amount of gas in galaxies. In principle, this is achieved by determining the HI mass function (HIMF), which is similar to the stellar mass function but specifically focuses on the neutral hydrogen content \citep{Baldry12}. While an indirect measurement of the HI cosmic density at redshift $z \sim 0.2$, has been measured using methods such as HI spectral stacking \citep{Delhaize13,Rhee18,Bera19,Chowdhury20}, Damped Ly-$\alpha$ absorption line systems \citep{Peroux03,Noterdaeme12,Grasha20}, or the [CII]-to-HI conversion factor \citep{Heintz21,Heintz22}. These studies generally agree that the HI mass density at low redshift undergoes minimal changes over time, unlike molecular hydrogen, which shows significant evolution and mirrors the overall rate of star formation. Other estimate on the HI cosmic density is shown in Table~\ref{tab:HI-cosmic-density-literature}. 

Beyond these existing observations, many experiments are underway to provide an unprecedented measurement of the HI signal in low redshift, median redshift, and high redshift universe. To this aim, several works have been done to predict the constraining power of the neutral hydrogen signal \citep{SantosMario15,StevenCunnington22,Karagiannis22,ZhangMing23}.
Forecasts have also been made on how well the HI IM joint survey strategy would constrain the dynamical dark energy and other cosmological parameters. For example, \citet{Elimboto19} predicted the improvement achieved by combining BINGO, FAST, and SKA-I telescopes with {\it Planck}.

Other work in this regard includes \citet{Zhang21}.
\citet{Wu23} considered a novel joint survey strategy using FAST $(0<z<0.35)$ + SKA1-MID $(0.35<z<0.8)$ + HIRAX $(0.8<z<2.5)$.

On the other hand, the thermal Sunyaev-Zel'dovich (tSZ) effect is a secondary anisotropy of the cosmic microwave background (CMB) produced by the inverse Compton scattering of CMB photons from warm-hot electrons in galaxy clusters \citep{carlstrom02}.
\begin{equation}
\label{eq:tsz}
\frac{\Delta T}{T_{\rm CMB}}= g(x)y,
\end{equation}
where $T_{\rm CMB}=2.725\,{\rm K}$ is the present day mean CMB temperature, $x\equiv h \nu/{ k_{\rm B}T_{\rm CMB}}$, with $h$ the Planck 
constant, $\nu$ the photon frequency and $k_{\rm B}$ the Boltzmann constant. The function $g(x)$ quantifies the frequency dependence of the tSZ effect, while the Compton parameter $y$ quantifies the amplitude of the tSZ effect independently of the observing frequency:
\begin{equation}
y = \Tilde{y} \int n_{\rm e}T_{\rm e}\,{\rm d}l,
\end{equation}
where $\Tilde{y} = {\sigma_{\rm T}k_{\rm B}}/{m_{\rm e}c^{2}}$, $T_{\rm e}$, $\sigma_{\rm T}$, and $n_{\rm e}$ are electron temperature, the Compton cross-section, and the electron number density respectively, and $m_{\rm e}$ is the electron mass. The tSZ effect is an integral one, taking contribution from all warm-hot gas from the surface of the last scatter to the observer. Much work has been done to model the electron pressure of the tSZ effect for low-mass range and high-mass range independently \citep{Battaglia12,Rotti21,Ibitoye22,Ibitoye23}. The observed tSZ is a good tracer of the large-scale structure and can provide insights into the gas properties, pressure profiles, and halo masses of galaxy clusters.

By cross-correlating the tSZ effect measured by Planck with the HI intensity mapping data, we may gain new insights into the process that governs the heating and cooling of gas in the Universe. The SZ effect is associated with the galaxy clusters, while at low redshift the HI gas is associated with galaxies. Such cross-correlation would help us understand the connection between galaxy clusters and the large-scale structure of the universe, and the HI content with the galaxy clusters.  It can probe the relation between the hot gas in galaxy clusters, and the underlying matter distribution traced by the HI signal, and shed light on the properties of galaxy clusters and their formation and evolution over cosmic time.  

The five-hundred-meter Aperture Spherical Telescope (FAST) \citep[FAST,][]{Nan11,Li_2016} is a single dish radio telescope that has the largest aperture (300 m during operation) to date and is equipped with 19 beams multibeam (L-band) feed system (future focal plane array >100), and low-noise cryogenic receivers. 
The large collecting area and high sensitivity make it a powerful tool for detecting neutral hydrogen in distant galaxies at low redshifts ($0.<z<0.35$). 

In this paper, we present for the first time a forecast of the cross-correlation signal between the tSZ effect and the HI intensity mapping field from FAST. 
A unique angle for the HI $\times$ tSZ signal is that it can be used (in the one-halo regime) to probe the neutral hydrogen abundance and profile within galaxy cluster/group halos, which is hard to probe in other traditional targeted observations.
To this aim, we employ a halo model framework to estimate the expected cross-correlation strength and explore its potential cosmological implications on the HI energy density parameter ${\Omega_{\rm HI}}$, the hydrostatic mass bias parameter ${1-b_{\rm H}}$ and other astrophysical parameters sensitive to the integrated electron-pressure of the tSZ. The tSZ effect is sensitive to the hot gas in galaxy clusters, and the cluster distribution traces the underlying dark matter. Similarly, in the post-reionization epoch, HI is all in galaxies that are tracing the underlying dark matter distribution too as well as the cluster distribution on large scales. Therefore, on large scales, both HI and tSZ are tracing the same distribution of matter and should be correlated to each other. This cross-correlating would take significant contributions from the low-redshift and high-mass objects in the Universe to probe the cosmic web and large-scale structure.

This paper is organized as follows. In Section~\ref{sec:HOD-model} we describe the theoretical prediction in a halo model framework. In 
Section~\ref{sec:results}, we present the methodology and results of our fisher matrix, 
and discuss the resultant implications. The concluding remarks are presented in Section~\ref{sec:conclusions}. 
Throughout this work we assume a spatially flat $\Lambda$-CDM cosmology with cosmological parameters 
fixed to {\it Planck\,} 2018 best-fitting values, i.e. $\Omega_{\rm c}h^{2}=0.120$, $\Omega_{\rm b}h^{2}=0.0223$, 
$h=0.674$, $n_{\rm s}=0.965$, $\tau=0.0540$ and $\ln(10^{10}A_{\rm s})=3.043$~\citep{planck2020}.

\section{HALO MODEL FORMALISM}
\label{sec:HOD-model}
The halo model is a method to describe the spatial distribution of matter, which works on the premise that all matter in the universe (e.g., galaxies) resides in virialized dark matter halos \citep{cooray2002}. This is applicable to neutral hydrogen as well, for example, \citet[]{villa14} found that almost all ($\gtrsim 90\%)$ of the {\rm HI}\ in the post-reionisation ($z < 6$) Universe resides within halos. 
There are two approaches to describe {\rm HI} in the halo model: the neutral hydrogen can be assumed to be simply distributed within each dark matter halo following a density profile (e.g., \citealt{Padmanabhan2016,Padman_Ref}); alternatively, one can model the distribution of galaxies within the halo using the halo occupation distribution (HOD) approach\citep{Berlind2002,Zheng2005}, then model the neutral hydrogen inside galaxies (e.g., \citealt{Laura19,Qin_2022}).

In the present work, we shall adopt the first approach, which ignores the distribution of discrete galaxies within a halo, but if we are mostly interested in scales above individual galaxies then this is sufficient, for the HI density profile within a halo may be regarded as a kind of average of the individual HI-harboring galaxies within a halo.

We proceed to compute the spatial distribution of HI using the model first proposed in \citet{Bagla10}
and subsequently employed in \citet{SKA_HI_fnl,Villaescusa15}, and \citet{Villaescusa16},
\begin{eqnarray}
M_{\rm HI}(M,z) = 
\begin{cases}
f(z)\,\,\frac{M}{1 + M/M_{\rm max}(z)} & {\rm if} M_{\rm min}(z) \leq M\\
0 & {\rm otherwise}
\end{cases} 
\end{eqnarray}
where the mass parameter $M_{\rm min}(z)$ and  $M_{\rm max}(z)$ represents the minimum mass of a dark matter halo able to host HI and the mass scale above which the fraction of neutral hydrogen in a halo becomes suppressed respectively. Their value corresponds to a dark matter halo with circular velocities, $v_{\rm circ} = 30\, {\rm km/s}$ (lower cut off) and $v_{\rm circ} = 200\, {\rm km/s}$, calculated using the virial relation
\begin{equation}
M = 10^{10} M_{\odot} \Bigg(\frac{v_{\rm circ}}{60 \,{\rm km/s}}\Bigg)^3 \Bigg( \frac{1 + z}{4}\Bigg)^{-3/2}.
\end{equation} 
The value of the free parameter $f(z)$ is chosen to reproduce the HI density parameter $\Omega_{\rm HI}(z)$, which is modeled as  $\Omega_{\rm HI} (z) = \Omega_{\rm HI}(1+ z)^{\alpha_{\rm HI}}$, where the value of $\alpha_{\rm HI}$ is set to its fiducial value in \citet{Villaescusa16}. 
The formalism adopted for the $M_{\rm HI}(M,z)$ function here, captures clusters, groups, and HI emitters that the HI $\times$ tSZ may be sensitive to (see Section 3.1 and 3.2 in \citet{villa14} for details of the difference between the two in modeling the spatial distribution of HI within halo). 

Inspired by observations \citep{Obreschkow09,Wang14} we model the HI halo profile  as:
\begin{equation}
\rho_{\rm HI}(r;M,z) = \rho_{0} \,{\rm exp} \left[-\frac{r}{r_{\rm s}(M,z)}\right],
\label{eq:profile}
\end{equation}
where $r_{\rm s}$ is a scale radius, defined here as $r_{\rm s}(M,z)\equiv R_{\rm v}(M)/c_{\rm HI}(M,z)$ \citep{Padman_Ref,Padmanabhan23}, $R_{\rm v}(M)$ stands for the virial radius of the dark matter halo of mass $M$, and $c_{\rm HI}$ denotes the concentration of the {\rm HI} comparable to the corresponding expression for dark matter, and is defined as \citep{maccio2007}
\begin{equation}
c_{\rm HI}(M,z) = c_{{\rm HI},0} \left(\frac{M}{10^{11} M_\odot} \right)^{-0.109} \frac{4}{(1+z)^\gamma}.
\end{equation} 
Analogous to the dark matter halo concentration parameter $c_{0} = 3.4$ in, e.g., \citet{maccio2007}, the typical value of the $c_{\rm HI,0}$ is usually higher than $c_{0}$ since HI gas is more tightly concentrated than the dark matter \citep{Padmanabhan2016}.
Note that $\rho_{0}$ in Equation~\ref{eq:profile}  is set by normalizing the {\rm HI} \, mass within the virial radius $R_{\rm v}(M)$, at a given halo mass $M$, and redshift $z$  to equal $M_{{\rm HI}}$

\begin{eqnarray} 
&& \int^{R_{\rm v}(M)}_{0} 4 \pi r^2 \rho_{{\rm HI}}(r) {\rm d}  r = M_{{\rm HI}}(M) \nonumber \\
\Rightarrow && \rho_0(M,z) = \frac{M_{\rm HI} (M)}{4\pi r_{\rm s}^3(M,z)f(c)}\label{eq:rho0}.
\end{eqnarray} 
where 
\begin{eqnarray}
f(c_{{\rm HI}}) 
&=& 2-{\rm e}^{-c}\left[2+c\left(c+2 \right)\right]. \label{eq:fc}
\end{eqnarray}

Below, we work with comoving coordinates. The Fourier transform of normalized {\rm HI} halo density profile $u_{\rm HI}(k|M,z)$ is given by
\begin{eqnarray}
\label{eq:u_HI}
u_{\rm HI}(k|M,z) &=& \frac{\int {\rm d}^{3}\mathbf{r}\,\exp\left(-i \mathbf{k}\cdot \mathbf{r} \right)\rho_{{\rm HI}}(ar|M,z)}{\int  {\rm d} ^{3}\mathbf{r}\,\rho_{{\rm HI}}(ar|M,z)} \nonumber \\
&=& \frac{1}{f(c_{{\rm HI}})}\frac{1}{k\tilde{r}_{\rm s}}u_{1}(k\tilde{r}_{\rm s}),
\end{eqnarray} 
where 
\begin{eqnarray}
\label{eq:u1}
&&u_{1}(k|M,z)= \frac{2(k\tilde{r}_{\rm s})}{\left(1+(k\tilde{r}_{\rm s})^2 \right)^{2}}\nonumber\\
&&-(k\tilde{r}_{\rm s})\frac{2+c(1+(k\tilde{r}_{\rm s})^{2})}{\left(1+(k\tilde{r}_{\rm s})^2 \right)^{2}}\cos(ck\tilde{r}_{\rm s}){\rm e}^{-c} \nonumber  \\
&&-\frac{1+c+(c-1)(k\tilde{r}_{\rm s})^2}{\left(1+(k\tilde{r}_{\rm s})^2 \right)^{2}}\sin(ck\tilde{r}_{\rm s}){\rm e}^{-c}.
\end{eqnarray}
where $\tilde{r}_{\rm s}=r_{\rm s}/a$ is the comoving scale radius. A derivation of Equation~\ref{eq:u1} is given in the Appendix. If we define the characteristic angular scale associated with the scale radius as $\ell_{\rm s}\equiv a\chi(z)/r_{\rm s}$, then $k\tilde{r}_{\rm s}\simeq \ell/\ell_{\rm s} \equiv \tilde{\ell}$, then Equation~(\ref{eq:u1}) can be expressed as
\begin{eqnarray}
\label{eq:u(l|M,z}
u_{1}(\tilde{\ell}|M,z) &=& \frac{2\tilde{\ell}}{\left(1+{\tilde{\ell}}^2 \right)^{2}}-(\tilde{\ell})\frac{2+c(1+{\tilde{\ell}}^{2})}{\left(1+{\tilde{\ell}}^2 \right)^{2}}\cos(c\tilde{\ell}){\rm e}^{-c} \nonumber  \\
&-&\frac{1+c+(c-1)\tilde{\ell}^2}{\left(1+{\tilde{\ell}}^2 \right)^{2}}\sin(c\tilde{\ell}){\rm e}^{-c}.
\end{eqnarray}
and ultimately Equation~(\ref{eq:u_HI}) can be summarized as
\begin{eqnarray}
u_{{\rm HI}}(\ell|M,z)=\frac{1}{f(c_{{\rm HI}})}\left(\frac{1}{\tilde{\ell}}\right)u_{1}\left( \tilde{\ell}| M,z\right).  
\end{eqnarray}

\subsection{Thermal SZ power spectrum}

For calculations involving the tSZ effect, it is convenient to define a ``3D Compton-$y$'' field that is simply a re-scaling of the electron pressure field:
\beq
\label{eq.3Dcomptonyfield}
y_{\rm 3D}(\vec{x}) = \frac{\sigma_{\rm T}}{m_{\rm e} c^2} P_{\rm e}(\vec{x}) \,.
\eeq
Note that the 3D Compton-$y$ field has dimensions of inverse length. 
The Fourier transform of the 3D Compton-$y$ profile for a halo of virial mass $M$ at redshift $z$ can be written as
\beqn
\tilde{y}_{\rm 3D}(\vec{k};M,z) & = & \int {\rm d}^3 r \, e^{-i \vec{k} \cdot \vec{r}} y_{\rm 3D}(\vec{r};M,z) \nonumber \\
				      & = & \int {\rm d}r \, 4 \pi r^2 \frac{\sin(kr)}{kr} y_{\rm 3D}(r;M,z) \,,
\label{eq.y3Dtwid}
\eeqn
In 2D, the SZ Fourier transform for a single halo of mass $M$ and redshift $z$ can then be computed as \citep{Planck2014_tSZ,Planck2016_sz,Ibitoye22}
\begin{eqnarray}
&&y_{\ell}(M_{500},z) =  \frac{4\pi R_{\rm 500}}{\ell^2_{\rm 500}} \frac{\sigma_{\rm T}}{m_{\rm e} c^2} \nonumber \\
	&&\times  \int {\rm d}x\, x^2 \frac{\sin(\ell x/\ell_{\rm 500})}{\ell x/\ell_{\rm 500}} P_{\rm e}(x;M_{500},z), \label{eq:y_ell}
\end{eqnarray} 
where we introduced the scaled radial separation $x=a\,r/R_{500}$ (with $a$ the scale factor
at the halo redshift) and $\ell_{\rm 500} =a \chi/R_{\rm 500}$. We evaluate the integral between the 
limits $x_{\rm min}=0$ and $x_{\rm max}=6$ as the physical scale $5\,R_{500}$ is usually considered marking the outer boundary of a galaxy cluster. We adopt the electron pressure profile parametrization 
derived in~\citet{Arnaud10}:
\begin{eqnarray}
\label{eq:pressprof}
&& P_{\rm e} \left(x;M_{500},z \right) 
=  1.65\,\, h^{2}_{70}\,\,  E^{8/3}(z)  \nonumber\\
&& \times   \left[ \frac{(1-b_{\rm H})M_{500}}{3 \times 10^{14}h^{-1}_{70} M_\odot} \right]^{2/3 + \alpha_{\rm p}} \mathbb{P}(x) \, {\rm [eV\,cm^{-3}]},  
\end{eqnarray} 
where $h_{70} = h/0.7$, 
$\alpha_{\rm p}\simeq 0.12$ represents the departure from the standard self-similar 
solution and $\mathbb{P}(x)$ is the ``universal'' pressure profile (UPP). The latter is parametrized as a generalized 
Navarro-Frenk-White profile~\citep{nagai07}:
\begin{equation}
	\label{eq:upp}
\mathbb{P}(x) = \frac{P_0}{(c_{500}x)^\gamma \left [1 + (c_{500}x)^\alpha   \right ]^{(\beta - \gamma)/\alpha}},
\end{equation} 
which we compute using the universal pressure profile (UPP) parameter values from~\citet{Planck2013}, 
$\left\{P_0,c_{500},\alpha,\beta,\gamma\right\}=\left\{6.41,1.81,1.33,4.13,0.31\right\}$.

\subsection{Cross-correlation angular power spectrum}

\begin{figure*}
	\centering
\centerline{\includegraphics[width=11cm]{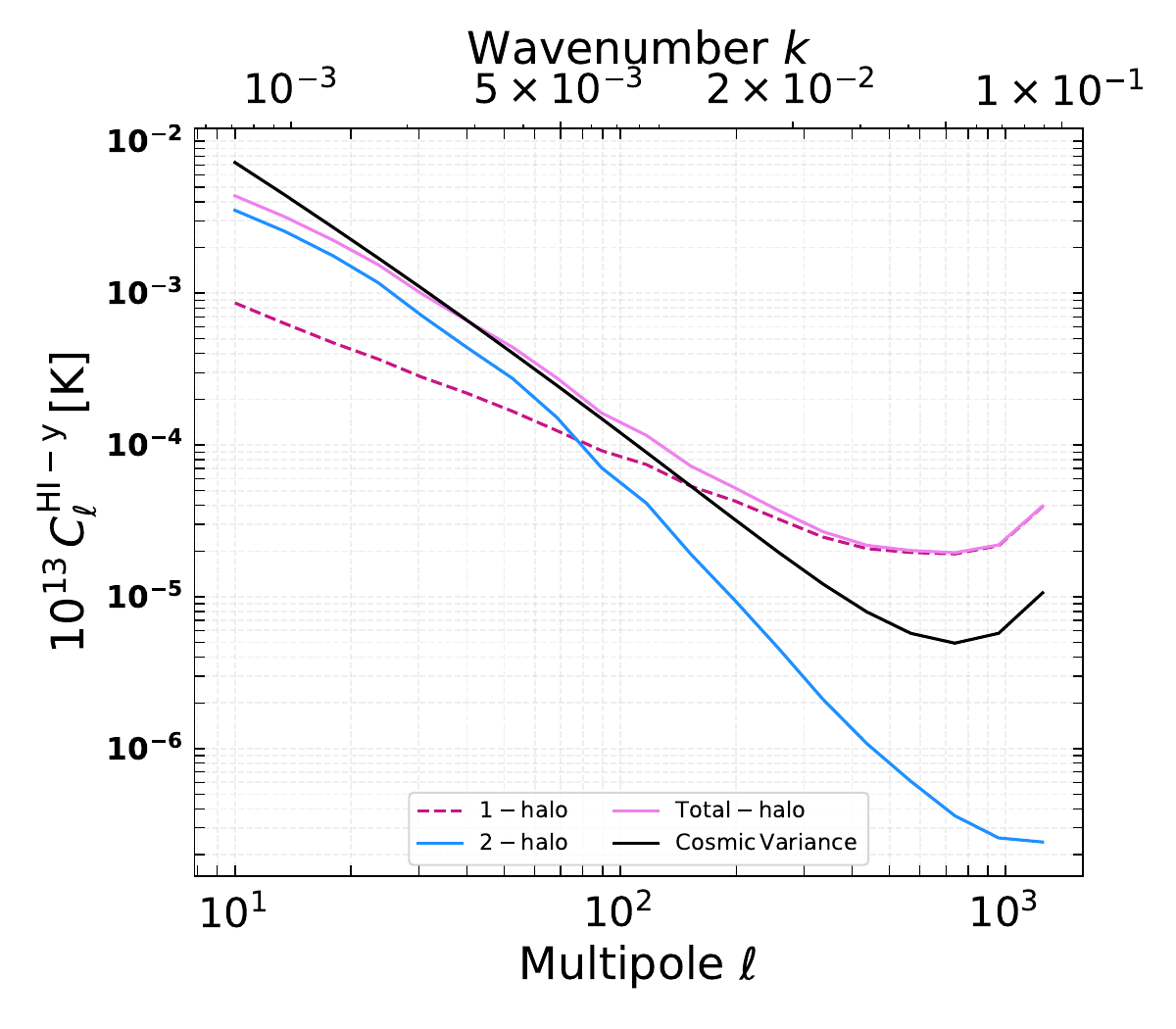}}
	\caption{The angular cross-correlation power spectra  $D_{\ell} \equiv \ell(1+\ell)\,C^{\rm HI-y}_{\ell}/2\pi$, the corresponding wave number $k$ is shown on the secondary x-axis. The total and the 1-halo and 2-halo terms are plotted. We also show the cosmic variance.}
	\label{fig:power-spectrum}
\end{figure*} 

The cross-correlation angular power spectrum can be expressed as the combination of 
the one-halo term and the two-halo term. The one-halo term is given by
\begin{eqnarray} 
\label{eq:1halo}
	&&C^{\rm XY,\text{1h}}_{\ell} = \int_{z_{\rm min}}^{z_{\rm max}} \textrm{d}z \frac{c \chi^2(z)}{H(z)}   \nonumber \\ &  \times &  \int^{M_{\rm max}}_{M_{\rm min}} \textrm{d}M \frac{{\rm d}n}{{\rm d}M}(M,z)\, 
X_{\ell}(M,z) Y_{\ell}(M,z),  
\end{eqnarray}
where $c \chi^2(z)/H(z) = {\rm d}^2V/({\rm d}z {\rm d}\Omega)$ is the comoving volume per unit redshift and solid angle, and $X_{\ell}$ and $Y_{\ell}$ are the spherical Fourier transforms of the corresponding generic observable on the sky. 

The two-halo term can be written as:
\begin{eqnarray}
\label{eq:2halo}
&&C^{\rm XY,\text{2h}}_{\ell} = \int_{z_{\rm min}}^{z_{\rm max}} \textrm{d}z \frac{c \chi^2(z)}{H(z)} P^{\rm{lin}}_{\rm{m}}\left(k = \frac{\ell +1/2}{\chi(z)},z\right) \nonumber \\
&&\times  \left[\int^{M_{\rm max}}_{M_{\rm min}}{\rm d}M\frac{{\rm d}n}{{\rm d}M}(M,z)\,b(M,z)X_{\ell}(M,z)\right] \nonumber \\
&&\times   \left[\int^{M_{\rm max}}_{M_{\rm min}}{\rm d}M\frac{{\rm d}n}{{\rm d}M}(M,z)\,b(M,z)Y_{\ell}(M,z)\right]. 
\end{eqnarray} 
we set $z_{\min}=10^{-3}$, $z_{\max}=5$ when 
computing the $yy$ auto-correlation,
as this interval safely includes all contributions from galaxy clusters. For the HI-$y$ cross-correlation 
and the HI auto-correlation, instead, we set the redshift according to the FAST instrumental parameters, 
i.e. $0< z< 0.35$. However, at $z_{\min}=0$ the computation of angular sizes diverges, 
we manually set the $z_{\min} =1\times10^{-3}$ to avoid it, in reality, this would not affect the result as SZ optical depth vanishes as $z\to 0$.  We set a lower mass limit of 
$10^{11}\,h^{-1}\text{M}_{\odot}$, below which the ICM pressure becomes negligibly low, and an upper limit 
of $10^{16}\,h^{-1}\text{M}_{\odot}$, after which the mass function severely cuts off the halo abundance. 

The measured quantity in HI intensity mapping is the temperature fluctuation per frequency slide, which can be written as~\citep{Battye2013}
\begin{eqnarray}
\delta T_{\rm b}(\hat{\mathbf{n}}) = \int {\rm d} \chi \,W_{{\rm HI}}(\chi)\bar{T}_{\rm b}(z)\delta_{{\rm HI}}\left(\chi\hat{\mathbf{n}}, z \right)
\label{eq:delta_Tb} 
\end{eqnarray}
where $\hat{\mathbf{n}}$ denotes sky direction, $W_{{\rm HI}}(\chi)$ is related to the top-hat projection kernel in redshift space by
$W_{{\rm HI}}(\chi) = W_{{\rm HI}}(z)~ { {\rm d}  z}/{ {\rm d}  \chi}$, with
\begin{eqnarray}
W_{{\rm HI}}(z)=
\begin{cases}
1/(z_{\rm max}-z_{\rm min}) & {\rm if}\,z_{\rm min}\leq z \leq z_{\rm max}\\
0 & {\rm otherwise}
\end{cases},    
\end{eqnarray}
$\bar{T}_{\rm b}(z)$ is the mean brightness temperature given by \citep{Hall2013,Battye2013}
\begin{eqnarray}
\label{eq:T-brightness}
\bar{T}_{\rm b}(z) &=& 0.18\,{\rm K}\,~\Omega_{{\rm HI}}h \frac{(1+z)^{2}}{E(z)}.     
\end{eqnarray}
Here we assume the local HI mass fraction $\Omega_{{\rm HI}}h=2.45\times 10^{-4}$ as given by the HI Parkes All Sky Survey (HIPASS)~\citep{Zwaan2005}. 
With
\begin{eqnarray}
\tilde{\delta}_{{\rm HI}}\left(k=\frac{\ell+1/2}{\chi(z)}; M,z \right)= \frac{M_{{\rm HI}}(M)}{\bar{\rho}_{{\rm HI}}(z)}\frac{\tilde{\ell}^{-1}}{f(c_{{\rm HI}})}u_{1}\left(\ell| M,z \right),\nonumber
\end{eqnarray}
the 2-D Fourier space, {\rm HI}\, temperature fluctuation field can then be calculated as~\citep{Hill2013}
\begin{equation}
\delta \tilde{T}_{\rm b}(\ell| M,z)  \simeq  
\bar{T}_{\rm b}(z)\frac{W_{\rm {\rm HI}}(\chi)}{\chi^{2}(z)}\frac{M_{{\rm HI}}(M)}{\bar{\rho}_{{\rm HI}}(z)}\frac{\tilde{\ell}^{-1}}{f(c_{{\rm HI}})}u_{1}\left(\ell| M,z \right).
\label{eq:Tb_ell}
\end{equation}

We can then substitute Equation~(\ref{eq:y_ell}) and (\ref{eq:Tb_ell}) as $A_{\ell}$ and $B_{\ell}$ in Equation~(\ref{eq:1halo}) and (\ref{eq:2halo}) to calculate the 1-halo and 2-halo terms. 

The ${\rm HI}-y$ cross-correlation angular power spectrum is shown in Figure~\ref{fig:power-spectrum}. On small scales ($\ell \gtrsim 70$), the 1-halo term dominates, while the 2-halo term, which follows the dark matter power spectrum, drops rapidly. On large scales, the 2-halo term contribute about 75\% of the total power spectrum. At these low \( \ell \) values, the contributions from cosmic variance are more pronounced due to the limited number of independent modes available in the observed volume of the universe. As $\ell$ increases, the contributions from shot noise and instrumental noise typically become more significant, reducing the impact of cosmic variance. The rise in amplitude at $\ell \gtrsim 10^3$ is introduced by the smoothing kernel.

\section{Fisher forecast}

\subsection{Fisher matrix formalism}
\label{ssec:Fisher_formalism}
We now use the Fisher matrix formalism to forecast the HI IM and its cross-correlation with the thermal SZ field. This is given by                           
\begin{equation}
\label{eq:Fisher_matrix}
F_{ab} =  \sum^{\ell_{\rm max}}_{\ell_{\rm min}}  \frac{1}{2} \, {\rm tr}[ C_{\ell, a} \,\Sigma _{\ell}\, C_{\ell, b}\,\Sigma _{\ell}],
\end{equation} where $\Sigma _{\ell}$ is the total noise inverse matrix,
\begin{equation}
    \Sigma _{\ell} = (N_{\ell})^{-1}.
\end{equation} 
The thermal noise for the single-dish intensity mapping experiment can be modeled as a Gaussian white noise with root-mean-square amplitude for each pixel given by \citep{Wilson}
\begin{equation}
\sigma_{\rm T} = \frac{T_{\rm sys}}{2\sqrt{t_{\rm pix} \delta v}},
\end{equation} 
where $T_{\rm sys}$ is the total system temperature, and $\delta v$ is the frequency channel bandwidth. If we have a  total observation time $t_{\rm obs}$, then the integration time per pixel is given by
\begin{equation}
t_{\rm pix} = t_{\rm obs} \frac{n_{\rm f}\Omega_{\rm pix}}{\Omega_{\rm survey}}, 
\end{equation}
where $n_{\rm f}$ is the number of feeds in the L-band, $\Omega_{\rm pix}$ is the pixel area which is equivalent to the square of the beam resolution, such that $\Omega_{\rm pix} \equiv \theta^2_{\rm FWHM}$.
The beam size of FAST is given by
\begin{eqnarray}
\theta_{\rm FWHM} &=& 1.22 \times \,\frac{21 {\rm cm} \,(1 + z)}{300\,{\rm m}}\nonumber \\
&=& 2.94 \,(1+z) \, {\rm arcmin}\nonumber
\end{eqnarray}
The noise angular power spectrum is given by
\begin{equation}
N^{\rm HI}_{\ell} = \frac{4 \pi}{N_{\rm pix}} {\sigma_{\rm T}}^2
\end{equation}

 \citet{Padmanabhan2020} (Equation 23) and  \citet{Lin22} (Equation 21), the variance of the cross-correlation angular power spectrum is given by
\begin{eqnarray}
\label{eq:HI-SZ noise}
{\rm N}^{\rm AB}_{\ell} =\frac{\delta_{{\ell}{\ell}}}{{f^{\rm AB}_{\rm sky}(2 \ell + 1)\Delta \ell}} \left[\hat{C}^{A}_{\ell}\,\, \hat{C}^{\rm B}_{\ell} + \hat{C}^{\rm AB}_{\ell}\hat{C}^{\rm AB}_{\ell} \right],
\end{eqnarray} 
where $\Delta \ell=1$, $A$, and $B$ are the observables, and the power spectra expressed as unit vectors for each observable is a sum of the power spectrum and thermal noise term in each case.

\subsection{Survey parameters}\label{sec:surveyparams}

We assume an HI intensity mapping survey with the FAST L-band multi-feed receiver system. Following, \citet{Elimboto21} we assume an average system temperature of $20\,{\rm K}$. We consider an HI survey with total observation time of $\sim 1\,{\rm yr}$, covering the full declination range allowed by FAST drift scan observation, $-14^\circ \sim 66^\circ$. The corresponding sky area is $\sim 24,000\ {\rm deg^2}$, which is also roughly the sky coverage of the Commensal Radio Astronomy FasT Survey \citep[CRAFTS,][]{Kai21}. However, the sky area within $\sim \pm 10\deg$ from the Galactic plane is heavily contaminated by the Galactic radiation, which is probably not usable, so we assume that the available survey area for the HI intensity mapping experiment is $\sim 20,000\ {\rm deg^2}$. 

The frequency range for the FAST HI survey is $1050  < \nu <  1420 {\rm MHz}$, 
corresponding to the redshift range of $0 \sim 0.35$.
However, the radio frequency interference (RFI)
can be a significant challenge for HI surveys.
\citet{FAST_drift_scan} noted that the FAST frequency band between $1150 {\rm MHz}$ to $1300 {\rm MHz}$ (see Figure 12 in \citet{FAST_drift_scan}) 
which corresponds to $0.09<z\leq0.235$ is strongly contaminated by the RFI produced by the Global Navigation Satellite Systems (GNSS). 
To this aim, we further break the integration in Equation \ref{eq:1halo}, and \ref{eq:2halo} 
into the redshift ranges corresponding to the two usable frequency bands, i.e.
the low-frequency band of $1050 < \nu < 1150 {\rm MHz}$ and
the high-frequency band is $1300 < \nu < 1420 {\rm MHz}$. 
The frequency resolution is set to $\Delta \nu = 10\,{\rm MHz}$ for cosmology studies. 
Thus, the low- and high-frequency bands are further binned into $10$ and $12$ frequency channels, respectively.

The FAST instrumental parameters and HI intensity mapping survey parameters assumed in this work
are summarized in Table~\ref{inst_and_survey_params}.

\begin{table} 
\begin{center}
\caption{FAST instrumental and survey parameters.}
\label{inst_and_survey_params}
\begin{tabular}{ll} 	
\hline\hline
Parameter Description & Value \\\hline 
\multicolumn{2}{c}{Instrumental Parameters$^\dag$} \\
Dish/aperture diameter & $D_{\rm dish} = 500$ m \\ 
Illuminated aperture diameter & $D=300$ m \\ 
Pointing accuracy      & $8$ arcsec \\ 
L-band frequency range, & $1,050 < \nu < 1,450$ MHz \\ 
Survey redshift range  & $0<z<0.35$ \\ 
System temperature     & $T_{\rm sys} = $ $20$ K \\ 
Number of L-band feeds & $n_{\rm f} = 19$ \\ 
L-band sensitivity     & $A_{\rm eff}/T_{\rm sys} = 1,600 \sim 2,000\,{\rm m}^{2}\,{\rm K}^{-1}$ \\ 
Telescope location     & ($25^{\circ}48'$ N, $07^{\circ} 21'$ E)\\  
FWHM at $1,420$ MHz    & $\theta_{\rm FWHM}=2.94$ arcmin \\ 
\hline
\multicolumn{2}{c}{Survey Parameters} \\
Frequency channel width      & $\Delta \nu = 10$ MHz  \\ 
\multirow[t]{2}{*}{Available frequency range}    & $1050$--$1150$MHz (low band) \\
& $1300$--$1420$MHz (high band)\\
Number of frequency channels & $N_{\nu} = 10 + 12$$^\ddag$\\
Sky coverage                 & $\Omega_{\rm survey} = 20,000\ {\rm deg}^{2}$ \\ 
Total observation time       & $t_{\rm obs} = 1\, {\rm yr}$ \\ 
Declination range            & $-14^{\circ}12' \sim 65^{\circ}48'$ \\ 
\hline\hline
\end{tabular}
\end{center}
$\dag$ \citet{Nan11, Li_2018, Jiang_2020}.\\
$\ddag$ For the low- and high-frequency band, respectively.
\end{table}

\subsection{Foreground}
\label{ssec:foreground_modeling}
The foregrounds primarily consist of diffuse synchrotrons and free-free emissions from both Galactic and extragalactic sources. These emissions are characterized as featureless and can be effectively removed by using smoothly varying functions, provided accurate bandpass calibration is available. Advanced foreground subtraction techniques have been developed specifically for dealing with intensity mapping (IM) data and similar data from experiments related to the Epoch of Reionization (EOR). These techniques have been developed and refined by various works, see e.g. \citet{Wang06,Liu11,Chapman13, Switzer13,Wolz17}.

The cross-correlation measurements are more robust against foregrounds than the 21cm auto-correlation measurements, as the other tracer is not affected by the foreground. Thus, for example, the previous forecasts of CMB-lensing $\times$ HI field ignored foregrounds \citep{Guha_Sarkar10,Dash21}. In the present work, we also ignore the influence of the foreground. 
In addition to this, \citet{Elimboto21} found that foreground removal does not have a significant effect on the auto-correlation of 21 cm for the FAST intensity mapping experiment. However, we restrict the sensitive $\ell$ -range in this study to $\ell \in [10,10^3]$,
as the cleaning of foregrounds from the HI intensity mapping effectively removes the largest scales at $\ell \leq 5$ \citep{Witzemann19,Cunnington19,Ballardini19}.
The high-$\ell$ cut is due to the choice of the resolution of the
tSZ map ($N_{\rm side}=512$).

The compton-y map of tSZ comes from the integral over wide redshift range \citep{Planck2016} and therefore mainly capture the large-scale radial flunctuations for each line-of-sight, i.e. the low $k_\parallel$ modes.
However, the low $k_\parallel$ modes for 21 cm signal are severely influenced by the Galactic foreground due to the smooth nature of the foreground. Therefore, the removal of the foreground for the 21 cm data may cause a severe loss of signal in the cross-correlation \citep{cunnington2023, roy2020}. 
Extracting the cross-correlation signal requires methods to recover the large-scale radial modes for 21 cm e.g. \citet{Cunnington19, wangzy2024} or data process methods that preserve more large-scale information, which is out of the scope of this paper.

\subsection{Effective Mass \& Redshift Range}
\label{sec:average_effective_mass}

\begin{figure}
\centering
\includegraphics[width=8cm]{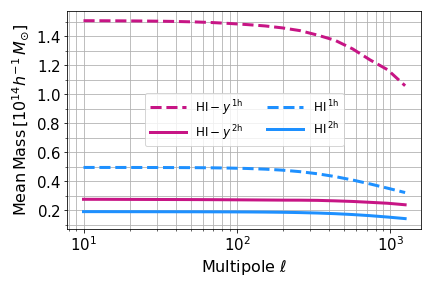}
\caption{The mean halo mass, as a function of $\ell$, which mostly contributes to the computation of the power spectra, calculated using Eq.~(\ref{eq:m1h}) for the 
	one-halo term and Eq.~(\ref{eq:m2h}) for two-halo term. Results are shown for all our correlations
	cases, marked as HI, HI-$y$.}
\label{fig:HI_and_HIy-mrange}
\end{figure}

The average mass of halos contributing to the 1-halo term in the cross-correlation can be written as \citep{Rotti21,Ibitoye22},
\begin{equation}
	\label{eq:m1h}	
	\langle M \rangle_{\ell}^{XY, {\rm 1h}} = \dfrac{\int^{z_{\rm max}}_{z_{\rm min}}\text{d}z \int^{M_{\rm max}}_{M_{\rm min}}\text{d}M \, M\, g_{\ell}^{XY}(M,z)}{ \int^{z_{\rm max}}_{z_{\rm min}}\text{d}z \int^{M_{\rm max}}_{M_{\rm min}}\text{d}M \,g_{\ell}^{XY}(M,z)},
\end{equation} 
where 
\begin{equation}
\label{eq:fab}
g_{\ell}^{XY}(M,z) = \dfrac{c\chi^2(z)}{H(z)}\,\dfrac{\text{d}n}{\text{d}M}(M,z)\,X_{\ell}(M,z)\,Y_{\ell}(M,z).
\end{equation}
For the 2-halo term,
\begin{equation}
	\label{eq:m2h}	
	\langle M^2 \rangle_{\ell}^{XY, {\rm 2h}} = \dfrac{\int^{z_{\rm max}}_{z_{\rm min}}\,\text{d}z\, w_{\ell}(z)\, Q(M \,b\,A_{\ell}) \, Q(M \,b\,B_{\ell}) } { \int^{z_{\rm max}}_{z_{\rm min}}\,\text{d}z\, w_{\ell}(z)\, Q(b\,A_{\ell})\, Q(b\,B_{\ell})},
\end{equation}
where
\begin{equation}
	w_{\ell}(z) = \dfrac{c\chi^2(z)}{H(z)}\,P^{\rm{lin}}_{\rm{m}}\left(\frac{\ell +1/2}{\chi(z)},z\right),
\end{equation}
and
\begin{equation}
	Q(x) =  \int^{M_{\rm max}}_{M_{\rm min}}\,\text{d}M \,x\, \dfrac{\text{d}n}{\text{d}M}(M,z).
\end{equation} 
The resulting average masses $\langle M \rangle_{\ell}^{XY, {\rm 1h}}$ and $\sqrt{\langle M^2 \rangle_{\ell}^{XY, {\rm 2h}}}$
for the ${\rm HI}$, ${\rm HI}y$ cases are
show in Figure~\ref{fig:HI_and_HIy-mrange}.

The halos contributing to the ${\rm HI}-y$ cross-correlation on large scale $(\ell \leq 200)$ have an average mass of a few $ 10^{14} M_{\odot}$ particularly in the one halo regime. We do expect that contribution to the Compton parameter $y$ should come from halos with masses $> 10^{14} M_{\odot}$. The 1-halo term receives contribution from more massive halos than the 2-halo term could also be expected. When $\ell > 200$, the average mass decreases, indicating that smaller halos start to make a notable impact on smaller scales. In terms of the HI auto-power contribution, we anticipate that it originates from significantly smaller halos. Indeed, on large scales, the average halo mass is around $\sim 10^{13} M_{\odot}$. This mass is substantially higher than that of typical galaxies, likely because, as mentioned in Section \ref{sec:HOD-model}, our model associates HI with halos rather than with individual galaxies.

\begin{figure}
\centering
\includegraphics[width=8cm]{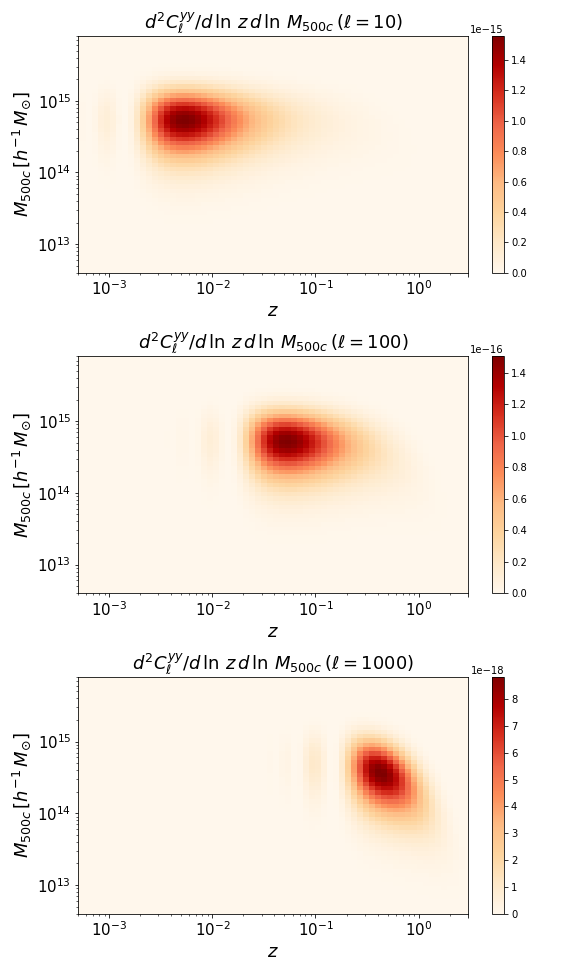}
\caption{Differential contributions with respect to halo mass and redshift for tSZ auto-power spectrum. In each row, the contributions for multipoles $\ell= 10, 100, 1000$ are shown. As expected, lower redshift and massive clusters contribute to the tSZ angular power spectrum at large scales (lower multipoles), and higher redshift, less massive objects set the tSZ amplitude at small scales (See Figure 5 and 6 in \citet{KomSel2002}}
\label{fig:yy-mrange}
\end{figure}

We can also study the differential contribution with respect to halo mass and redshift to the $yy$ auto power spectrum \citep{Ken-Osato2019},
\beqa
\frac{{\rm d}^2 C^{yy} (\ell)}{{\rm d}z{\rm d}M} &=&
\frac{{\rm d}^2 C^{yy}_\mathrm{1h} (\ell)}{{\rm d}z{\rm d}M} +
\frac{{\rm d}^2 C^{yy}_\mathrm{2h} (\ell)}{{\rm d}z{\rm d}M}, \\
\frac{{\rm d}^2 C^{yy}_\mathrm{1h} (\ell)}{{\rm d}z{\rm d}M} &=&
\frac{{\rm d}^2V}{{\rm d}z d\Omega} \frac{{\rm d}n (M,z)}{{\rm d}M}
|\tilde{y}_\ell (M,z)|^2 , \\
\frac{{\rm d}^2 C^{yy}_\mathrm{2h} (\ell)}{{\rm d}z{\rm d}M} &=&
\frac{{\rm d}^2V}{{\rm d}z d\Omega} P_\mathrm{m} \left( k=\frac{\ell+1/2}{D_A (z)}, z \right)
\nonumber \\
&& \times 2 \left[ \frac{{\rm d}n (M, z)}{{\rm d}M}
\tilde{y}_\ell (M, z) b_\mathrm{h} (M, z) \right.
\nonumber \\
&& \left. \int \!\! {\rm d}M \frac{{\rm d}n (M, z)}{{\rm d}M}
\tilde{y}_\ell (M, z) b_\mathrm{h} (M, z) \right] .
\eeqa
This is shown in Figure~\ref{fig:yy-mrange}, for three multipoles $\ell = 10, 100, 1000$. For the higher $\ell$, the contribution comes from higher redshift $z$, and for the low $\ell$ which corresponds to the large angular scale, most contributions come from the low redshift. The masses of the halo which makes contribution do not differ much. As illustrated for the $\ell=10 $ case, 
in the top panel of Figure~\ref{fig:yy-mrange},  relatively massive and low redshift objects that contribute significantly to the $yy$-auto power spectrum would better correlate with the neutral hydrogen signal from FAST, whose redshift is $0<z<0.35$.

One can see that the HI-auto and HI-$y$ spectra are sensitive to a wide range of objects, including galaxies of different masses, while the tSZ signal is driven by more massive objects. This is because while the neutral hydrogen signal is representative of the overall distribution of neutral hydrogen in galaxies of various masses, the tSZ signal primarily reflects the distribution and properties of galaxy clusters, which are massive objects. 

In this work, we have restricted our forecast to scales above clusters, i.e. $\ell \leq 1500$ where the massive halos with more HI could also have a stronger tSZ effect. We expect the HI and tSZ to be correlated at these scales. At even smaller scales (the tSZ effect can only reach a scale of $\ell \leq 3000-4000$), the neutral gas resides in small halos with less ionized gas, there may be some anti-correlation between the two probes. In addition, at this scale, we may need to consider the distribution of the gas within the halo more carefully. The HI resides in central and satellite galaxies, which deviates from the continuous HI distribution assumed here. This would happen on scales that typically correspond to multipoles $\ell \geq 3 \times 10^4$ or equivalently to modes with wave numbers of the order of $k \geq 2 h/\,{\rm Mpc}$ \citep{Laura19} (see Appendix~\ref{sec:HI-continuous or discrete} for a comparison between the two HI distribution). We leave the investigation of the small-scale anti-correlation to future work. 

\section{RESULTS}
\label{sec:results}

We forecast the constraints on cosmological and astrophysical parameters in a $\Lambda$CDM paradigm. 
The key cosmological parameter to which our anticipated cross-correlation power spectrum is particularly sensitive is the cosmic density of neutral hydrogen $\Omega_{\rm HI}$, while the astrophysical parameters encompass the hydrostatic mass bias parameter $B_{h}$ and the universal pressure profile (UPP) parameters (${P_0},c_{500},\alpha_y,\beta_y$). Indeed, the cosmic density of the HI can be estimated by the observation of the HI emission in the local universe. To forecast the measurement expectation of the HI cosmic density from the cross-correlation of FAST HI and Planck tSZ observation, we present the results obtained by combining the Fisher matrix information from the $yy$-auto correlation, ${\rm HI}$-autocorrelation and the ${\rm HI}y$-cross correlation. Consequently, in this section, we illustrate the marginalized one- and two-dimensional distributions of the parameters of interest, as depicted in Figure~\ref{fig:HI-y_constraints}. As a verification step, we equally discussed the expected constraints in the absence of the ${\rm HI}$-auto-correlation.

\subsection{Constraints on $\Omega_{\rm HI}$}   
\label{sec:Discuss_HI_constraints}

\begin{table*}
\centering	
\caption{
Comparison between different constraints from 21cm-related science, including the constraints on HI cosmic density, expressed as $\Omega_{\rm HI}$. Constraints are broadly grouped into observational and
theoretical/simulation. Observational constraints include those from galaxy surveys, DLA observations, and HI intensity mapping experiments. For each case, we report the main
technique employed in the analysis, the parameter(s) constrained, the corresponding mean
redshift/redshift range, where available, and the reference in the literature for each.}
\label{tab:HI-cosmic-density-literature}

\scalebox{0.8}{
\begin{tabular}{llll} 
\hline\hline
Technique   & Constraints ( $\Omega_{\rm HI}$ are $h^{-1} \times 10^{-4}$) & Mean redshift  & Reference\\
&&(Redshift range)&\\ \hline
\multicolumn{4}{c}{Theory/Simulation}\\
\textbf{FAST HI Cross-correlation forecast} & $  \mathbf{\sigma}(\Omega_{\rm \textbf{HI}})= \,\,\mathbf{0.01} $ & $ \mathbf{0 - 0.35}$ & \textbf{This work}\\ \textbf{with tSZ} \\
 The HI from BINGO project combined & $\Omega_{\rm HI} = (6.2 \pm 4.1) h$ &   $ 0.127 - 0.449$ & \citet{Costa_BING22}  \\ with {\tt Planck}\\
{Hydrodynamical simulation using } & $\Omega_{\rm HI} = (1.4 \pm 0.18) h$ & 0 &\\
GADGET-2/OWLs & $\Omega_{\rm HI} = (2.5 \pm 0.14) h$ & 1 & \citet{Duffy12} \\
   &  $\Omega_{\rm HI} = (3.8 \pm 0.08) h$ & 2  & \\
                  
{Galaxy formation simulation, HI} & $\Omega_{\rm HI} = 4.3 \pm 0.3$ & $\sim 0.1$ & \\ 
 astrophysics, based on SKA-MDB2 & $\Omega_{\rm HI} = 4.60 \pm 1.0$ & $\sim 1.0$ & \citet{Chen21}\\ 
 Survey and SKA-DB1 Survey\\ 
 
{N-body simulation, HI prescription}
& $\Omega_{\rm HI} = (11.2 \pm 3.0) h$ & $\sim 0.8$  &\citet{Khandai11} \\
 combined with \citet{first_HI_science}\\
\hline
\multicolumn{4}{c}{DLA observations} \\     

{DLA measurements from HST and } & $\Omega_{\rm HI} = 5.2 \pm 1.9$    & 0.609 (0.11 - 0.90)  &  \\
SDSS & $\Omega_{\rm HI} = 5.1 \pm 1.5$    & 1.219 (0.90 - 1.65) & \citet{Rao06}\\
& $\Omega_{\rm HI} = 4.29^{+0.24}_{-0.23}$ & (2.2 - 5.5) & \citet{Prochaska09} \\
 & $\Omega_{\rm HI}(z)$ & (2.0 - 5.19)  & \citet{Noterdaeme09, Noterdaeme12} \\ 
Cross-correlation of DLA and Ly-$\alpha$ & $b_{\rm DLA} = 2.17 \pm 0.2 $ &  $\sim 2.3$ &\citet{Font-Ribera12} \\
 forest observations\\
Observations of DLAs with HST/COS & $\Omega_{\rm HI} = 9.8^{+9.1}_{-4.9}$ & $< 0.35$ & \citet{Meiring11} \\
DLAs and sub-DLAs with VLT/UVES & $\Omega_{\rm HI}(z)$ & 1.5 - 5.0 & \citet{Zafar13} \\

\hline
\multicolumn{4}{c}{HI intensity mapping} \\ 
{WSRT HI emission}  & $\Omega_{\rm HI} = 2.22 \pm 0.40$ & 0.1 & \\
                        & $\Omega_{\rm HI} = 2.29 \pm 0.61$ & 0.2 & \citet{Rhee13}\\ 

{DINGO HI emission} & $\Omega_{\rm HI} = 4.20 \pm 0.8$  & 0.057 & \\
                        &  $\Omega_{\rm HI} = 4.60 \pm 0.7$ & 0.008 & \citet{Rhee23}\\ 

Cross-correlation of DEEP2 galaxy-HI & $\Omega_{\rm HI} b_{\rm HI} r^{\dagger} = (5.5 \pm 1.5)h$ & 0.8 & \citet{first_HI_science} \\
 fields\\
HI intensity fluctuation  &  $\Omega_{\rm HI} b_{\rm HI} r = (4.3 \pm 1.1)h$  & 0.8 & \citet{Masui13} \\
cross-correlation with WiggleZ survey\\

MeerKAT HI IM pilot survey & $\Omega_{\rm HI} b_{\rm HI} r = (8.6 {\pm 1.0}\,(\textrm{stat}) \pm{1.2}\,(\textrm{sys}))$  & $0.4- 0.459$ & \citet{Cunnington23} \\      
 cross-correlation with WiggleZ survey\\
HI auto-power spectrum combined with &  $\Omega_{\rm HI} b_{\rm HI} = 6.2^{+2.3}_{-1.5}h$ & $0.8$ &  \citet{Switzer13}\\ 
 cross-correlation with WiggleZ survey\\
\hline
\multicolumn{4}{c}{Galaxy surveys} \\

ALFALFA HI emission  & $\Omega_{\rm HI}^{*} = 3.0 \pm 0.2$  & 0.026 & \citet{Martin10} \\ 

HIPASS HI emission   & $\Omega_{\rm HI} = 2.6 \pm 0.3$      & 0.015 & \citet{Zwaan2005}\\ 

{HIPASS, Parkes; HI stacking} 
& $\Omega_{\rm HI} = 2.82^{+0.30}_{-0.59}$ & 0.028 (0 - 0.04) &{\citet{Delhaize13}} \\
& $\Omega_{\rm HI} = 3.19^{+0.43}_{-0.59}$ &  0.096 (0.04 - 0.13) & \\

AUDS HI emission from galaxies +  HI & $\Omega_{\rm HI} =  2.63 \pm 0.1$ & 0.065 (0.0 - 0.2) & \citet{Hoppmann15} \\
 surveys + stacking \\

GMRT HI emission stacking & $\Omega_{\rm HI} = (5.0 \pm 1.8)h$ & 0.32  & \citet{Rhee18} \\ 

uGMRT HI emission stacking  & $\Omega_{\rm HI} =  (4.81 \pm 0.75)h$ & $0.2<z<0.4$ & \cite{Bera19} \\ 

uGMRT HI stacking, DEEP2 field & $\Omega_{\rm HI} =  (4.5 \pm 1.1)h$ & $ \sim 1.06$ & \cite{Chowdhury20} \\ 

MIGHTEE-HI: first MeerKAT HI mass function & $\Omega_{\rm HI} =   5.46^{+0.94}_{-0.99}$ & $ 0 \leq z \leq 0.084$ & \cite{Ponomareva23} \\ 

 MIGHTEE-HI: first MeerKAT HI mass  & $\Omega_{\rm HI} =   6.31\pm {0.31}$ & $ 0 \leq z \leq 0.084$ & \cite{Ponomareva23} \\
 function obtained using Modified &&&\\
 Maximum Likelihood (MML)\\
 
HI distribution maps from M31, M33  & $\Omega_{\rm HI} = 3.83 \pm 0.64$ & 0.0 & \citet{Braun12} \\ 
and LMC\\

\hline
$\dagger$ $r$ denotes the stochasticity. & & &

\end{tabular} 

}
\end{table*}

The Fisher matrix takes as input the power spectra computed with the halo model framework, to obtain
\begin{eqnarray}
    \sigma(\Omega_{\rm HI}) = 1.0 \times 10^{-6}.
\end{eqnarray} 
In \citet{Switzer13,Masui13} and \citet{Cunnington23} a joint constraint is placed on the product $b_{\rm HI} \Omega_{\rm HI} r$, where $b_{\rm HI}$ is the bias and $r$ is the galaxy–hydrogen correlation coefficient ($r$). Here the HI bias is calculated independently by using the halo model, and hence the constraint is on the HI cosmic density,  in contrast to the model adopted  If the HI does not completely trace the halo distribution, then there may be a correlation factor $r < 1$, but here we omit this factor. This result is comparable to those obtained in \citet{Deng22}, though here we adopted a different fiducial cosmic HI density value, $\Omega_{\rm HI}=2.5 \times 10^{-4}$, while in \citet{Deng22} a higher fiducial value of $\Omega_{\rm HI}=5 \times 10^{-4}$ was used. This constraint appears smaller than the previous estimate on the error of the HI cosmic density because we have assumed a larger sky coverage $\sim 20, 000 {\rm deg^2}$ and longer integration time as explained in Section~\ref{sec:surveyparams}. 
Here we also included auto-correlation in making the constraint. Without the auto-correlation, the error on $\Omega_{\rm HI}$ increased substantially by $>$ 40\%. The reduced precision with which the HI cosmic density can be determined in the absence of this constraint highlights the significance of the HI-auto correlation. Hence several works are ongoing to obtain useful constraints on the 21cm auto power spectrum to date \citep{Paul23,Moodley23,Elahi24}, but even if the auto power is not detected with the precision given by the forecast due to the complicated foreground, the constraint obtained from the cross power is still quite stringent. 

As shown in Figure \ref{fig:HI-y_constraints},
we find an anti-correlation between the HI concentration parameter quantified by $c_{\rm HI,0}$ and the cosmic neutral hydrogen relative density $\Omega_{\rm HI}$. A future work that explore this correlation may be useful to investigate the physical processes that may lead to this degeneracy.

\subsection{Constraints on the UPP parameters}   

\begin{figure*}
	\centering
    \includegraphics[width=7.3in]{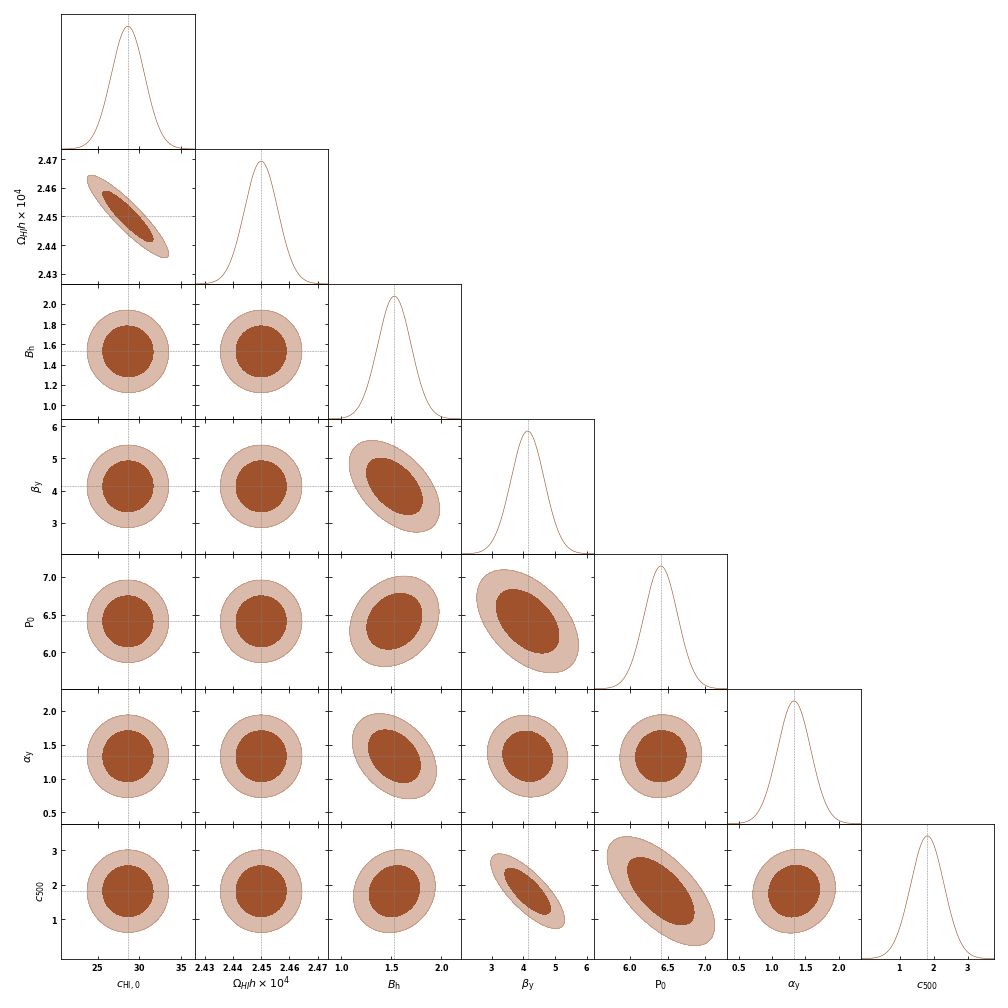}
\caption{Marginalized distributions (diagonal) and 2D correlations (off-diagonal) plots of the six independent parameters considered in the halo model framework. The dashed lines indicate the input parameter values.}
	\label{fig:HI-y_constraints}
\end{figure*}

\label{sec:UPP-parameters-constraints}

\begin{table*} 
\centering
\caption{Comparison between different constraints for the parameters defining the universal pressure profile (UPP) of clusters. For each case we give the reference in the literature, the main physical observable(s) employed in the analysis, the number of objects used in the study, and the best-fit values for the UPP parameters. Boldface values were kept fixed in the corresponding fit.  ``WL'' refers to weak lensing for brevity.}
\label{tab:upp_summary}
\renewcommand{\arraystretch}{1.3}
\begin{tabular}{cccccccc}

\hline
\multirow{4}{*}{\textbf{Reference}} & \multicolumn{2}{c}{\multirow{2}{*}{\textbf{Data set}}} & \multicolumn{4}{c}{\multirow{2}{*}{\textbf{UPP parameters}}} \\
 & & & & & & & \\
\cline{2-8}
 & \multirow{2}{*}{\textbf{Objects}} & \multirow{2}{*}{\textbf{Observables}} & \multirow{2}{*}{$\mathbf{P_0}$} & \multirow{2}{*}{$\mathbf{c_{500}}$} & \multirow{2}{*}{$\mathbf{\alpha_y}$} & \multirow{2}{*}{$\mathbf{\beta_y}$} \\
 & & & & & & & \\
\hline\\
\textbf{This work} & - & \textbf{SZ, FAST}  & $\mathbf{6.41\pm{0.22}} $ & $ \mathbf{1.81\pm 0.48} $ & $\mathbf{1.33 \pm 0.24} $ & $ \mathbf{4.13 \pm 0.52}$\\[1ex]
 & & & & & & & \\
\multirow{2}{3.4cm}{\centering \citet{nagai07}} & \multirow{2}{1.8cm}{\centering 16 clusters} & \multirow{2}{2.3cm}{\centering X-ray, simulations} & \multirow{2}{1.8cm}{\centering 3.3} & \multirow{2}{1.4cm}{\centering 1.8} & \multirow{2}{1.4cm}{\centering 1.3} & \multirow{2}{1.5cm}{\centering 4.3}\\
 & & & & & & & \\
\multirow{2}{3.4cm}{\centering \citet{Arnaud10}} & \multirow{2}{1.8cm}{\centering 33 clusters} & \multirow{2}{2.3cm}{\centering X-ray, simulations} & \multirow{2}{1.8cm}{\centering 8.403$\,h_{70}^{-3/2}$} & \multirow{2}{1.4cm}{\centering 1.177} & \multirow{2}{1.4cm}{\centering 1.0510} & \multirow{2}{1.5cm}{\centering 5.4905} \\
 & & & & & & & \\
\multirow{2}{3.4cm}{\centering \citet{Planck2013}} & \multirow{2}{1.8cm}{\centering 62 clusters} & \multirow{2}{2.3cm}{\centering SZ, X-ray} & \multirow{2}{1.8cm}{\centering 6.41} & \multirow{2}{1.4cm}{\centering 1.81} & \multirow{2}{1.4cm}{\centering 1.33} & \multirow{2}{1.5cm}{\centering 4.13}\\[1ex]
 & & & & & & & \\
\citet{Sayers16} & 47 clusters & SZ, X-ray & $9.13\pm2.98$ & $\mathbf{1.18}$ & $\mathbf{1.0510}$ & $6.13\pm0.76$\\[1ex]
\citet{Gong19}  & $\sim10^5$ LRGs & SZ & $2.18^{+9.02}_{-1.98}$ & $1.05^{+1.27}_{-0.47}$ & $1.52^{+1.47}_{-0.58}$ & $3.91^{+0.87}_{-0.44}$\\[1ex]
\multirow{2}{3.4cm}{\centering \citet{Ma21_convergence}} & \multirow{2}{1.8cm}{\centering -} & \multirow{2}{2.3cm}{\centering SZ, WL (convergence)} & \multirow{2}{1.8cm}{\centering $9.68^{+10.02}_{-7.11}$} & \multirow{2}{1.4cm}{\centering $2.71^{+0.92}_{-0.93}$} & \multirow{2}{1.4cm}{\centering $5.97^{+1.81}_{-4.73}$} & \multirow{2}{1.5cm}{\centering $3.47^{+1.39}_{-0.60}$} \\[1ex]
 & & & & & & & \\
\citet{Ma21_convergence} & - & SZ, WL (shear)  & $6.62^{+2.06}_{-1.65}$ & $1.91^{+1.07}_{-0.65}$ & $1.65^{+0.74}_{-0.50}$ & $4.88^{+1.18}_{-2.46}$\\[1ex]

\multirow{2}{3.4cm}{\centering \citet{Pointecouteau21}} & \multirow{2}{1.8cm}{\centering 31 clusters} & \multirow{2}{2.3cm}{\centering SZ} & \multirow{2}{1.8cm}{\centering $3.36^{+0.90}_{-0.71}$} & \multirow{2}{1.4cm}{\centering $\mathbf{1.18}$} & \multirow{2}{1.4cm}{\centering $1.08^{+0.13}_{-0.11}$} & \multirow{2}{1.5cm}{\centering $4.30\pm0.12$}\\
 & & & & & & & \\
 
\multirow{2}{3.4cm}{\centering \citet{He21}} & \multirow{2}{1.8cm}{\centering 33 clusters} & \multirow{2}{2.3cm}{\centering X-ray, simulations} & \multirow{2}{1.8cm}{\centering 5.048} & \multirow{2}{1.4cm}{\centering 1.217} & \multirow{2}{1.4cm}{\centering 1.192} & \multirow{2}{1.5cm}{\centering $\mathbf{5.490}$}\\
 & & & & & & & \\
 
\multirow{2}{3.4cm}{\centering \citet{Tramonte23}} & \multirow{2}{1.8cm}{\centering $\sim 2.3 \times 10^{4}$ clusters} & \multirow{2}{2.3cm}{\centering SZ} & \multirow{2}{1.8cm}{\centering $5.9^{+2.3}_{-2.0}$} & \multirow{2}{1.4cm}{\centering $2.0 \pm 0.7$} & \multirow{2}{1.4cm}{\centering $ 1.8 \pm 0.7 $} & \multirow{2}{1.5cm}{\centering $4.9^{+1.2}_{-1.0}$}\\[1ex]
 & & & & & & & \\[0.3ex]

\multirow{2}{3.4cm}{\centering \citet{Melin23}} & \multirow{2}{1.8cm}{\centering 31 clusters} & \multirow{2}{2.3cm}{\centering SZ} & \multirow{2}{1.8cm}{\centering $3.36^{+0.90}_{-0.71}$} & \multirow{2}{1.4cm}{\centering $\mathbf{1.18}$} & \multirow{2}{1.4cm}{\centering $1.08^{+0.13}_{-0.11}$} & \multirow{2}{1.5cm}{\centering $4.30\pm0.12$}\\
 & & & & & & & \\ 
\hline
\end{tabular}
\end{table*}

The pressure profile parameters provide valuable insights into the formation of galaxy clusters, the nature of dark matter, and dark energy, and the measurement of cosmological parameters such as matter density and the expansion rate of the universe. By understanding the pressure distribution within the galaxy clusters, we can gain insight into their growth and the underlying physical processes involved. 
For the simulation data with fiducial parameters,
\begin{eqnarray}
    {P_0},c_{500},\alpha_y,\beta_y = [0.22, 0.48, 0.24, 0.52].
\end{eqnarray} 
we obtain constraints. This and the constraints in the literature are shown in Table~\ref{tab:upp_summary}.

It may be useful to explain the implications of our results on the relationship between normalization $P_{0}$ and other parameters. It determines the overall amplitude of the pressure profile and sets the level or magnitude of the pressure at a specific reference radius, often chosen to be the characteristic scale radius $R_{\rm s}$.  As shown in Figure~\ref{fig:HI-y_constraints}, we see that larger values of the hydrostatic mass bias parameter prefer a slightly higher pressure profile amplitude ($P_{0}$ is correlated with $B_{\rm h}$). This may arise due to various physical processes, such as gravitational collapse. The collapse of matter under gravity may tend to concentrate towards the center of the system, which in turn leads to the enhancement of both pressure and matter density, resulting in a correlation between the $P_{0}$, and $B_{\rm h}$. On the other hand, $P_{0}$ is basically uncorrelated with $\alpha_y$, but anti-correlation with the concentration parameter $C_{500}$. While the concentration parameter characterizes the steepness or compactness of a density profile within a system, the normalization factor sets the level of the pressure profile in amplitude or intensity. Thus, in systems with low concentration, the density reduces because mass is spread out over the large volume, which in turn may lead to a lower value of $P_{0}$. 
Similarly, we can see an anti-correlation between the slope parameter that characterizes the radial dependence of the pressure profile, $\beta_\gamma$, and the concentration parameter that describes the concentration of the gas distribution within the system, $C_{500}$. This degeneracy may be easily understandable as different combinations of the two parameters are needed to maintain the hydrostatic equilibrium (e.g For a steeper profile, a higher value of $\beta$ requires a lower $c_{500}$ and the opposite for a shallower profile). This result is consistent with recent stacking analysis using the Planck catalog. See the left plot of Figure 12 in \citet{Tramonte23}.

For other further comparison with the literature, we summarize and compare our results with other measurements of the UPP parameters in Table \ref{tab:upp_summary}. Our results show the ability of this cross-correlation analysis to contribute to the understanding of the distribution of gas pressure within galaxy clusters.

\section{Conclusions} 
\label{sec:conclusions}
Cosmology with 21cm-neutral hydrogen intensity mapping is a rapidly evolving field, thanks to improvements in observations. Some useful measurements at specific redshifts have already been obtained  (see Table.~\ref{tab:HI-cosmic-density-literature} for a summary of previous constraints), and upcoming experiments with better signal-to-noise ratio, higher resolution, and larger sky fraction would provide a more precise constraint on the HI signal. In the meantime, it is important to forecast their impact on cosmology and more importantly, what constraints are possible from their cross-correlation with other large-scale structure probes. In this work, therefore, we have studied the capability to detect the underlying HI signal from the cross-correlation measurement of the FAST HI IM map with the thermal SZ map on large scales based on a halo model framework. 
We considered the implication of the cross-correlation power spectrum on the cosmic neutral hydrogen density parameter. In the presence of instrumental noise and the absence of foreground, we forecast
\begin{eqnarray}
    \sigma_{\rm HI} = 1.0 \times 10^{-6}.
\end{eqnarray}   
could be achieved. The error of the HI concentration parameter $c_{\rm HI,0}$ is estimated as 3.7. A result that is smaller compared to the error estimate preferred by \citet{Padmanabhan2016} where a redshift evolution of the HI concentration parameter was adopted.
We also demonstrate that this cross-correlation is useful for studying the astrophysical implications of the universal pressure profile parameters. 
We obtained error estimates (standard deviations) on the parameters as
\begin{eqnarray}
    {B_{h}},{P_0},c_{500},\alpha_y,\beta_y = [0.16,0.22, 0.48, 0.24, 0.52].
\end{eqnarray} 
It is important to note that the correlation between different parameters of the pressure profile is determined by the physical processes at play in the system as detailed in Section~\ref{sec:UPP-parameters-constraints} . It is possible that certain physical mechanisms could lead to an anticorrelation between some parameters. For example, if there is a dominant process that affects the central region of the cluster, it could lead to an inverse relationship between the central pressure and the slope of the pressure profile. It is also possible that the parameters are not strongly correlated or that the correlation depends on other factors. The precise relationship between pressure profile parameters can vary depending on factors such as the cluster's mass, redshift, or specific physical processes that impact the distribution of gas. For an example, see the constraints at different masses and redshift in Figure 19 and Figure 21 in \citet{Tramonte23}. We present a comparison of our error on the UPP parameters with other analyses in Table~\ref{tab:upp_summary}.

Many avenues remain for future work on this topic. Our work focuses more on the theoretical forecast and does not consider scales beyond $\ell \sim 1500$, meaning that we could not test possible systematics and also do not explore the impact of this cross-correlation on smaller scales. Analysis incorporating tests for possible systematics such as those arising from the survey area effect, foreground effect, RFI flagging, and noise effect would likely lead to more accurate cosmological constraints from the HI-tSZ angular power spectrum.
In addition, a complementary study of the joint analysis of HI-auto, tSZ-auto, galaxy-auto and the cross-correlation between the three observables combined in a $6 \times 2$ pt analyses modeling and inference can help break parameter degeneracies and enhance the precision of cosmological measurements. Although the HI $\times$ tSZ, HI $\times$ HI, tSZ $\times$ tSZ, HI $\times$ galaxy and tSZ $\times$ galaxy measurements may all be sensitive to the distribution of matter, their expected gain is however unique. 
The HI $\times$ tSZ offer insights into the spatial distribution of neutral hydrogen and the hot gas within galaxy clusters traced by the tSZ effect. We think that a $6 \times 2$ pt cross-correlation of such a combination of observables may offer a complementary probe of the large-scale structure of the universe.

Finally, we have shown the potential of the cross-correlation power spectrum between FAST HI and tSZ to constrain cosmological and astrophysical parameters at relatively low redshift. Observational constraints of the neutral hydrogen density with minimal foreground challenge can be obtained with enhanced calibration techniques and more observing time from ongoing experiments of both intensity mapping and galaxy surveys \citep{FAST_drift_scan}, and SKA \citep{SKA-1_HI_gal_survey,Cunnington23}. 

Similarly, future and upcoming experiments, such as the Cosmic Microwave Background Stage-4 experiments (CMB-S4; \citealt{Abazajian19, Abazajian22, Barron18}) and the Atacama Cosmology Telescope (ACT; \citealt{Thornton16, Ramasawmy22}) would provide high-resolution y maps of  about $0.5 \mu$K-arcmin,15 arcseconds at
150 GHz \citep{Sehgal19} and $1.6^{\prime}$ \citep{Coulton24} respectively. While, the former is expected to cover $\sim$ 70\% of the entire sky, the latter would cover the southern and equatorial regions and would obtained a better constraints on small scales when cross-correlated with the FAST HI data covering the northern sky. The South Pole Telescope (SPT; \citealt{Bleem22}) on the other hand would provide a higher resolution of a lowest-noise Compton y-map with angular resolution of $1.25^{\prime}$ and would provide a better capture of the correlation on small scales with HI data covering the same sky.

\section*{Acknowledgements} 
We would like to thank Oluwayimika Ibitoye, Professor Huanyuan Shan, Professor Amare Abebe and Akinlolu Ibitoye for helpful discussions. AI acknowledges the support of the National SKA Program of China with grant No. 2022SKA0110100 and the Alliance of International Science Organizations Visiting Fellowship on Mega-Science Facilities for Early-Career Scientists (Grant No: ANSO-VF-2022-01, and ANSO-VF-2024-01). YZM acknowledges the support from the National Research Foundation of South Africa with Grant No. 150580, No. 159044, No. CHN22111069370 and No. ERC23040389081. YG acknowledges the support from the National Key R\&D Program of China grant Nos. 2022YFF0503404, 2022SKA0110402, and the CAS Project for Young Scientists in Basic Research (No YSBR-092). XL acknowledges the support of the Ministry of Science, and Technology (MoST) inter-government cooperation program China-South Africa Cooperation Flagship Project 2018YFE0120800

\section*{Software and data}

For the analysis presented in this manuscript, we made use of the following software: {\sc GetDist}~\citep{lewis19}.
The data underlying this article is publicly available. The \textit{Planck} Compton 
parameter noise map is available at the {\it Planck} Legacy Archive at. 
\url{http://pla.esac.esa.int/pla}.

\appendix

\section{Choice of HI distribution}
\label{sec:HI-continuous or discrete}
It is important to note that there are two different treatments of the HI distribution as detailed in \citet{Laura19} which are the continuous HI distribution and the discrete HI distribution. For a continuous HI distribution, the spatial distribution of neutral hydrogen in a dark matter halo may be traced using an independent density profile such as the NFW or cored-NFW profile which does not require a detailed modeling of the central and satellite galaxy contribution (for instance, see \citet{Padman_Ref,Padmanabhan23}). This is suitable to model the gas distribution at the end of the reionization epoch and the early stages of galaxy formation. On the other hand, a discrete HI distribution takes into account the halo contribution from all HI-emitting objects and hence treats HI in central galaxies differently from HI in satellite galaxies. The two are consistent to quantify how the HI traces the dark matter distribution up to wave number $k \leq 2 h/\,{\rm Mpc}$ (equivalent to $\ell \sim 3 \times 10^4$) if the discrete HI include a scale-independent Poisson noise contribution but extends to $k \leq 10 h/\,{\rm Mpc}$ ($\ell \sim 1.4 \times 10^5$) if it does not contain a scale-independent Poisson noise contribution (See Figure 4 of \citet{Laura19}). To this aim, we conclude that the HI distribution in our work is sufficient for the scale we are interested in.


\bibliography{Reference}{}
\bibliographystyle{aasjournal}

\label{lastpage}
\end{document}